\documentclass[letterpaper,12pt]{article}
\usepackage{color,amssymb,enumerate,float}

\usepackage{ifpdf}
\ifpdf
	\usepackage[pdftex]{graphicx}
	\usepackage[pdftex,unicode,implicit]{hyperref}

	\hypersetup{
  	pdftitle     = {}, 
  	pdfkeywords  = {},
  	pdfauthor    = {},
  	pdfcreator   = {pdf\LaTeXe\ with package \flqq hyperref\frqq},
  	pdfproducer  = {pdf\LaTeXe\ with package \flqq hyperref\frqq},
  	pdfpagemode  = UseNone,  
  	pdffitwindow = true,  
  	unicode      = true,
  	plainpages   = true,
  	colorlinks   = true,  
  	citecolor    = black,  
  	urlcolor     = blue, 
  	linkcolor    = black
	}

\else

  \usepackage[dvips]{graphicx}

\fi 

\makeatletter
\@addtoreset{equation}{section}
\makeatother

\begin{document}
	
\thispagestyle{empty}

\begin{center}

{\bf \LARGE On the space of solutions of the Ho\v{r}ava theory at the kinetic-conformal point}
\vspace*{15mm}

{\large Jorge Bellor\'{\i}n}$^{a,1}$
{\large and Alvaro Restuccia}$^{a,b,2}$
\vspace{3ex}

{\it $^a$Department of Physics, Universidad de Antofagasta, 1240000 Antofagasta, Chile.}
{\it $^b$Department of Physics, Universidad Sim\'on Bol\'{\i}var, 1080-A Caracas, Venezuela.} 
\vspace{3ex}

$^1${\tt jbellori@gmail.com,} \hspace{1em}
$^2${\tt arestu@usb.ve}

\vspace*{15mm}
{\bf Abstract}
\begin{quotation}{\small  
 The nonprojectable Ho\v{r}ava theory at the kinetic-conformal point is defined by setting a specific value of the coupling constant of the kinetic term of the Lagrangian. This formulation has two additional second class-constraints that eliminate the extra mode. We show that the space of solutions of this theory in the Hamiltonian formalism is bigger than the space of solutions in the original Lagrangian formalism. In the Hamiltonian formalism there are certain configurations for the Lagrange multupliers that lead to solutions that cannot be found in the original Lagrangian formulation. We show specific examples in vacuum and with a source. The solution with the source has homogeneous and isotropic spatial hypersurfaces. The enhancement of the space of solutions leaves the possibility that new solutions applicable to cosmology, or to other physical systems, can be found in the Hamiltonian formalism.	
}
   \end{quotation}

\end{center}

\thispagestyle{empty}

\newpage
\section{Introduction}

Ho\v{r}ava theory \cite{Horava:2009uw} has been studied as a candidate for a perturbatively renormalizable theory of quantum gravity. Its main characteristic is that it abandons the symmetry of general covariance over the spacetime in order to introduce terms of higher order in spatial derivatives in the Lagrangian. These terms can improve the renormalization of the theory. In this scenario, the emerging of undesirable ghosts can, in principle, be under better control than in generally covariant theories \cite{Stelle:1976gc}, since the order in time derivatives is not increased.

The first step in the formulation of the theory is the assumption of the existence of a foliation constituted of spatial slices, each one corresponding to a instant of time. The underlying symmetry of the theory is given by the diffeomorphisms that preserve the foliation (FDiff). The FDiff do not allow that the transformation of the time coordinate depend on the spatial points; the congruence of lines of time is preserved by the FDiff. One implication of this reduced symmetry is that a spacetime structure is not mandatory \cite{Horava:2009uw}. Although the theory is formulated in terms of the standard Arnowitt-Deser-Misner (ADM) variables, the geometry is encoded on the spatial metric, whereas the lapse function and the shift vector can be regarded as fields over the spatial slices that evolve in time. A spacetime structure can be anyway implemented by interpreting the lapse function and the shift vector as the corresponding components of the spacetime metric, but this is not mandatory.

The nonprojectable version of the Ho\v{r}ava theory \cite{Horava:2009uw,Blas:2009qj}, which is the case defined by the possibility of the lapse function depends on the spatial coordinates, admits a special formulation that is given by setting a particular value of the coupling constant of the kinetic term of the Lagrangian, which is $\lambda = 1/3$. When this value is set, the theory acquires two additional second-class constraints and the extra mode is eliminated. This case was studied in terms of its Hamiltonian formulation in Ref.~\cite{Bellorin:2013zbp}. Interestingly, at the value $\lambda = 1/3$ the kinetic term of the Lagrangian acquires an anisotropic conformal symmetry \cite{Horava:2009uw}, but the theory is not conformally invariant since the potential is not. For this reason in previous works we have called this case as the kinetic-conformal formulation. In Ref.~\cite{Bellorin:2016wsl} we showed that the kinetic-conformal theory has propagators only for the transverse traceless tensorial modes of the spatial metric (on the transverse gauge), and that there are no ghosts. We also proved the power-counting renormalizability of the theory in that reference. 

A recent report on the Ho\v{r}ava theory can be found in \cite{Wang:2017brl}. We emphazise that in the kinetic-conformal theory there are no radiative corrections to the $\lambda = 1/3$ condition. This is not associated to symmetries, it is a dynamical fact. The second-class constraints are not associated to gauge symmetries, the elimination of the extra mode in the kinetic-conformal formulation is not due to a gauge symmetry. The quantization is based on the constraints and field equations, they must be satisfied at any stage of the process of quantization. Indeed, the two additional second-class constraints of the kinetic-conformal theory are incorporated to the path integral and affect the measure \cite{Bellorin:2016wsl}. The coupling constants of the theory are effectively modified by the raditive corrections, but this must happen in such a way that the constraints are preserverd, otherwise the quantization is inconsistent. Therefore, the constraints of the theory necessarily fix $\lambda = 1/3$, $\lambda$ is not a running constant in the kinetic-conformal formulation. The quantum kinetic-conformal formulation is an independent theory with its own degrees of freedom, it is not a smooth limit of the Ho\v{r}ava theory with $\lambda \neq 1/3$.

Our aim in this paper is to point out that in the kinetic-conformal theory there are more classical solutions in the Hamiltonian formulation than in the original Lagrangian formulation. A key condition leading to this behavior is that the canonical formulation requires the addition of all the constraints, primary and secondary, to the Hamiltonian. In other words, although the Hamiltonian formulation can be obtained from the Lagrangian one by the standard Legendre transformation, the resulting Hamiltonian, once all the constraints have been added to it, looses the exact equivalence with the original Lagrangian. Actually, the classification of constraints between primary and secondary is a consequence of the Lagrangian taken to obtain the primary Hamiltonian \cite{dirac}. When the quantization of the theory is intended, the correct Hamiltonian is the one with all the constraints added.

We give concrete examples of the enhancement of the space of solutions by studying vacuum (i.~e., purely gravitational) configurations and configurations with matter sources. In the first case we find examples with static solutions. In the second case we incorporate a homogeneous and isotropic perfect fluid. We present a solution that has homogeneous and isotropic spatial slices with a scale factor for the spatial metric that is time dependent. We present these solutions mainly to illustrate the point about the additional solutions in the Hamiltonian formulation.

In the Lagrangian formulation of the kinetic-conformal Ho\v{r}ava theory there is a strong restriction on a class of homogeneous and isotropic configurations. If all the ADM variables and the source, which is taken as a perfect fluid, are assumed to be homogenous and isotropic, and the perfect fluid is assumed to adopt the same relativistic form of general relativity (GR), then the only possibility allowed by the Lagrangian field equations when $\lambda = 1/3$ is that the density and the pressure vanish (for the flat case, specifically). Actually, in Ref.~\cite{Blas:2009qj}, where the extension of the nonprojectable theory and its first applications were presented, this result arises rather as a divergence on the cosmological gravitational constant. This is so because in that paper the field equations were managed to obtain a deformation of the Friedman equations valid, in principle, for any value of $\lambda$. In this way, in the Friedmann equations there arises a rescaled cosmological gravitational constant, which has a pole at $\lambda = 1/3$. But, as we have commented, the actual interpretation of this class of configurations at $\lambda = 1/3$ is that the source is empty. The same restriction can be seen in the Einstein-aether theory \cite{Jacobson:2000xp}, if the analog of the kinetic-conformal condition is imposed (in terms of the coupling constants of the Einstein-aether theory). This occurs already in the general case of the Einstein-aether theory, for which the deformation of the Friedmann equations was established in \cite{Mattingly:2001yd,Carroll:2004ai,Jacobson:2008aj}. For the particular case of the aether vector is restricted to be hypersurface orthogonal, which is called the khronometric theory or the $T$-theory, and it is the case equivalent to the Lagrangian of the nonprojectable Ho\v{r}ava theory truncated at second order in derivatives \cite{Blas:2009ck,Jacobson:2010mx}, an analysis on its cosmological configurations was applied in \cite{Jacobson:2013xta,Audren:2014hza}. In this hypersurface orthogonal case there arises the same restriction on the homogeneous and isotropic configurations when the kinetic-conformal condition is imposed.

We remark that all these studies about cosmological backgrounds are based on the Lagrangian formulation.  According to the analysis we present here, in the Hamiltonian formulation there is still open the possibility of finding more solutions with the features of homogeneity and isotropy. We also remark that the Hamiltonian  formulation we study here applies to the Ho\v{r}ava theory; we do not pretend to extend this formulation to the Einstein-aether theory. Another important assumption that deserves discussion is the condition of homogeneity and isotropy on all the ADM variables. In Ho\v{r}ava theory the structure of a spacetime is not mandatory since it is not a generally covariant theory, as we have commented. Strictly speaking, the principle of homogeneity and isotropy is restricted to the spatial metric, whereas the lapse function and the shift vector are fields over the spatial slices. In this scenario, it is admisible to consider lapse functions or shift vectors that can be inhomogeneous or anisotropic, but being accompanied by a homogeneous and isotropic spatial metric, whenever the field equations admit such configurations as solutions. We shall ellaborate on this point, in particular by relaxing the conditions on the lapse function. 

We comment that there are several solutions known for the (complete) nonprojectable Ho\v{r}ava theory. Most of them have been obtained in the effective theory truncated at second order in derivatives. Indeed, some of them are known from (or simultaneosly discovered in) the Einstein-aether theory, owing on the equivalence both theories share at the level of the Lagrangians (for the case in which the solutions meet the condition of hypersurface orthogonality). Static spherically symmetric wormholes (and naked singularities) were found in the Einstein-aether in Ref.~\cite{Eling:2006df}. The same solutions were recovered on the side of the nonprojectable Ho\v{r}ava theory in Refs.~\cite{Kiritsis:2009vz,Bellorin:2014qca}. An extension of these wormholes for the case of negative cosmological constant was done in \cite{Bellorin:2015oja}. Black holes that are also static and spherically symmetric\footnote{In Ho\v{r}ava theory, the vacuum spherically symmetric solutions do not possess the uniqueness properties that they have in GR.} have been found in Refs.~\cite{Barausse:2011pu,Blas:2011ni}. These black holes possess universal horizons, despite the fact that the underlying theory is not relativistic. For further discussion on the physics of universal horizons see, for example, \cite{Barausse:2013nwa,Cropp:2013sea,Lin:2014ija,various:universalhorizon}. Slowing rotating black holes were found in Refs.~\cite{Barausse:2012qh,Wang:2012nv}. Recently an aspect of the thermodynamics associated to the universal horizon has been studied in Ref.~\cite{Pacilio:2017emh}, specifically a quite general method for deriving the Smarr formula for the Ho\v{r}ava theory and the Einstein-aether theory, including the case of slowly rotating rotating black holes. There are also many important solutions for the projectable theory and for some truncated models of the nonprojectable theory (imposing the detailed balance principle, for example), as well as in lower dimensional models. For example, a recent model in $1+1$ dimensions of the Ho\v{r}ava theory coupled to a nonrelativistic scalar field has been studied in Ref.~\cite{Li:2017aow}, finding black holes with universal horizons.

This paper is organized as follows: to illustrate on simple grounds how in some systems there can be differences on the space of solutions of the Hamiltonian and the Lagrangian, we discuss in section \ref{sec:toymodel} a simple toy model that mimics the gravitational theory and that exhibits this behavior. In section \ref{sec:generalhamiltonian} we summarize the complete Hamiltonian formalism of the kinetic-conformal nonprojectable Ho\v{r}ava theory. In section \ref{sec:vacuumstatic} we discuss new vacuum solutions in this formalism. In section \ref{sec:cosmosolution} we present a homogeneous and isotropic solution with a source. In section \ref{sec:lagrangian} we check explicitly that the new solutions cannot be found in the original Lagrangian. Finally, we present some conclusions.


\section{Hamiltonian vs Lagrangian equations in a toy model}
\label{sec:toymodel}
Let us start with the standard approach of building the Hamiltonian of a theory by performing a Legendre transformation on a given Lagrangian. Consider a field theory over one time direction and one spatial direction and whose independent fields are $\phi(t,x)$ and $n(t,x)$. The idea with this model is that $\phi$ plays the role of analog of the spatial metric and $n$ the one of the lapse function. We denote by a dot the time derivative and by a prime the spatial derivative, $\dot{\phi} = \partial \phi / \partial t$, $\phi' = \partial \phi / \partial x$. We define the original Lagrangian such that the action is
\begin{equation}
 S = 
 \int dt dx \, n \left[ 
   \frac{1}{2} \left( \frac{ \dot{\phi} }{ n } \right)^2 - V 
 - \left( \frac{ n' }{ n } \right)^2  \right] \,.
 \label{actionlagrangian}
\end{equation}
$V = V(\phi,\phi',\ldots)$ is a potential for $\phi$ that depends on it and its spatial derivatives, but not on the time derivatives, and does not depend on $n$. The last term in (\ref{actionlagrangian}) is the potential for $n$ and there is no kinetic term for it. In this toy model we fix the values of all the coupling constants for the sake of simplicity. The equations of motion of this action are
\begin{eqnarray}
   \partial_t \left( \frac{ \dot{\phi} }{ n } \right) 
 + \frac{ \delta \left< n V \right> }{ \delta \phi } &=& 0 \,,
 \label{eomlagrangian1}
 \\
 \frac{1}{2} \left( \frac{ \dot{\phi} }{ n } \right)^2 
 + V 
 + \left( \frac{ n' }{ n } \right)^2 - \frac{ 2 n'' }{ n } &=& 0 \,,
 \label{eomlagrangian2}
\end{eqnarray}
where the brackets $\left<\,\,\right>$ denote total integration, $\left< f \right> = \int dt dx f$.

Let us perform a Legendre transformation on the Lagrangian given in (\ref{actionlagrangian}). We denote by $\pi$ the momentum conjugated to $\phi$, which results
\begin{equation}
 \pi = \frac{ \dot{\phi} }{ n } \,,
 \label{legendretransform}
\end{equation}
and by $p_n$ the momentum conjugated to $n$, which is zero. This is the unique primary constraint of the theory. The Hamiltonian density obtained in this way is
\begin{equation}
 H^{(0)} = 
 n \left[ 
  \frac{1}{2} \pi^2 + V + \left( \frac{ n' }{ n } \right)^2 \right]  
 + \sigma p_n \,,
 \label{hamiltonianprimitive}
\end{equation}
where we have incorporated the primary constraint with the Lagrange multiplier $\sigma$. 

We now apply the Dirac procedure for the time preservation of the constraints. The preservation of the primary constraint yields the secondary constraint
\begin{equation}
 \mathcal{H} \equiv
 \frac{1}{2} \pi^2 + V 
 + \left( \frac{ n' }{ n } \right)^2 - \frac{ 2 n''}{n}  = 0 \,,
 \label{hamiltonianconstraint}
\end{equation}
which we call the ``Hamiltonian" constraint. $p_n$ and $\mathcal{H}$ are second-class constraints, since
\begin{equation}
 \{ \mathcal{H}(t,x) , p_n(t,y) \} =
   \frac{ 2 }{ n } \left[ \frac{ n'' }{ n } 
 - \left( \frac{ n' }{ n } \right)^2 \right]
 \delta( x - y ) 
 - 2 \left( \frac{ \delta(x-y) }{ n } \right)''
 \,.
 \label{brackethp}
\end{equation}
The other combinations of Poisson brackets are zero, $ \{ \mathcal{H} , \mathcal{H} \} = \{ p_n , p_n \} = 0$. Consequently, the preservation of the Hamiltonian constraint yields an equation for the Lagrange multiplier $\sigma$, namely
\begin{equation}
   \sigma''  - n' \left( \frac{ \sigma }{ n } \right)'
 - \frac{ n'' \sigma }{ n }  = 0 \,.
 \label{sigmaeq}
\end{equation}
With Eq.~(\ref{sigmaeq}) Dirac's procedure ends; the constraints of the theory are $p_n$ and $\mathcal{H}$.

The Hamiltonian $H^{(0)}$ given in (\ref{hamiltonianprimitive}) includes only the primary constraint. As we have commented, the classification between primary and secondary constraints is subject to the choice of the Lagrangian made in (\ref{actionlagrangian}) \cite{dirac}. Therefore, we define the total Hamiltonian as
\begin{equation}
 H =  n \left[ 
 \frac{1}{2} \pi^2 + V 
 + \left( \frac{ n' }{ n }\right)^2 \right]  
      + \sigma p_n + A \mathcal{H} \,,
 \label{hamiltoniantotal}
\end{equation}
where $A$ is a Lagrange multiplier. Actually, constraint $p_n = 0$ is already solved. Since it plays no role in the rest of our analysis, we substitute the condition $p_n = 0$ explicitly in the Hamiltonian for the sake of simplicity. We define the total canonical action as
\begin{equation}
 S_{\mbox{\tiny CAN}} =
 \int d^2x \left[ 
   \pi \dot{\phi}  - n \left( 
    \frac{1}{2} \pi^2 + V + \left( \frac{ n' }{ n } \right)^2 \right)
 - A \mathcal{H} \right]   \,.
 \label{actioncanonical}
\end{equation}

Now we derive the independent equations of motion of the action (\ref{actioncanonical}) by taking variations of it with respect to $n$, $\pi$, $\phi$ and $A$. Variations with respect to $A$ yield the Hamiltonian constraint $\mathcal{H} = 0$ given in (\ref{hamiltonianconstraint}). Variations with respect to $n$, after imposing $\mathcal{H} = 0$ explicitly, yield the equation\footnote{In the case of this toy model the Eqs.~(\ref{sigmaeq}) for $\sigma$ and (\ref{aeq}) for $A$ result to be the same, but even in this case the solutions for $\sigma$ and $A$ do not need to be the same.}
\begin{equation}
 \left( \frac{ A }{ n } \right)''
 + \frac{ n' }{ n } \left( \frac{ A }{ n } \right)' \,
 = 0 \,.
 \label{aeq}
\end{equation}
We may present the formal general solution of this equation in terms of an indefinite integral of $n^{-1}$ and two arbitrary functions of time $k_1(t)$ and $k_2(t)$,
\begin{equation}
 \frac{A}{n} = k_1 \int \frac{ dx }{ n } + k_2 \,.
 \label{solutionaeq}
\end{equation}
Variations with respect to $\pi$ and $\phi$ yield, respectively,
\begin{eqnarray}
 \dot{\phi} &=& ( n + A ) \pi \,,
 \label{dotphi}
 \\
 \dot{\pi} &=& - \frac{ \delta \left< ( n + A ) V \right> }{ \delta \phi} \,.
 \label{dotpi}
\end{eqnarray}
 
In summary, to get a classical solution of the theory defined by the total canonical action (\ref{actioncanonical}), one must find a set of fields $\phi$, $\pi$, $n$ and $A$ that solve the Eqs.~(\ref{hamiltonianconstraint}), (\ref{aeq}), (\ref{dotphi}) and (\ref{dotpi}). The solutions of (\ref{aeq}) can always be deduced from (\ref{solutionaeq}).

Now we proceed to show that the space of solutions of the canonical action (\ref{actioncanonical}) is bigger than the space of solutions of the Lagrangian action, whose equations of motion are (\ref{eomlagrangian1}) and (\ref{eomlagrangian2}). First, the canonical equations imply the Lagrangian ones. To see this, we assume that
\begin{equation}
 n + A \neq 0 \,,
 \label{nonsingular}
\end{equation}
such that the relation between $\phi$ and $\pi$ in Eq.~(\ref{dotphi}) can be inverted,
\begin{equation}
 \pi = \frac{ \dot{\phi} }{ n + A } \,.
\end{equation}
By substituting this relation in the Hamiltonian constraint (\ref{hamiltonianconstraint}) and the Eq.~(\ref{dotpi}), we get the two equations
\begin{eqnarray}
    \frac{1}{2} \left( \frac{ \dot{\phi} }{ n + A } \right)^2 
  + V + \left( \frac{ n' }{ n } \right)^2 - 2 \frac{n''}{n}  &=& 0 \,,
  \label{preksquare}
  \\
    \partial_t \left( \frac{ \dot{\phi} }{ n + A } \right) 
  + \frac{ \delta \left< ( n + A ) V \right> }{ \delta \phi} &=& 0 \,.
  \label{predotphi}  
\end{eqnarray}
Now, from the general solution (\ref{solutionaeq}) we extract the class of solutions given by $k_1(t) = 0$ and $k_2(t)$ arbitrary, such that $A = k_2(t) n$, requiring only $k_2(t) \neq -1$ in order to meet condition (\ref{nonsingular}). On this class of configurations the Eqs.~(\ref{preksquare}) and (\ref{predotphi}) become
\begin{eqnarray}
\frac{1}{2} \left( \frac{ \dot{\phi} }{ (1 + k_2) n } \right)^2 
+ V + \left( \frac{ n' }{ n } \right)^2 -  \frac{2 n''}{n}  &=& 0 \,,
\label{fromhtol1}
\\
\partial_t \left( \frac{ \dot{\phi} }{ (1 + k_2) n } \right) 
+ \frac{ \delta \left< ( (1 + k_2) n ) V \right> }{ \delta \phi} &=& 0 \,.
\label{fromhtol2}
\end{eqnarray}
It is evident that the redefinition of $n$ given by $n \rightarrow (1 + k_2) n$ renders this two equations, which remain as the only unsolved equations on the side of the canonical variables, identical to the Lagrangian equations (\ref{eomlagrangian1}) and (\ref{eomlagrangian2}).

The second part is to show that there are more solutions on the canonical side. If we now assume
\begin{equation}
 n + A = 0 \,,
 \label{singular}
\end{equation}
which is obtained from (\ref{solutionaeq}) by putting $k_1(t) = 0$ and $k_2(t) = -1$, we get that relation (\ref{dotphi}) cannot be inverted to solve $\pi$. Hence, the connection with the original Lagrangian by the way of a Legendre transformation is lost. Instead, we get that the Eqs.~(\ref{dotphi}) and (\ref{dotpi}) become
\begin{equation}
 \dot{\phi} = \dot{\pi} = 0 \,.
\end{equation}
The Hamiltonian constraint (\ref{hamiltonianconstraint}) remains as a condition on $\phi$, $\pi$ and $n$. Therefore, this class of solutions is given by $A = -n$, static fields $\phi$ and $\pi$, and a field $n$. The last three variables must satisfy constraint (\ref{hamiltonianconstraint}), which is the only condition left on them.

To complete the comparison of this class of solutions with the Lagrangian formulation, we substitute a static field $\phi$ in the Eqs.~(\ref{eomlagrangian1}) and (\ref{eomlagrangian2}), obtainning
\begin{eqnarray}
\frac{ \delta \left< n V \right> }{ \delta \phi } &=& 0 \,,
\label{extra}
\\
 V 
+ \left( \frac{ n' }{ n } \right)^2 - \frac{ 2 n'' }{ n }  &=& 0 \,.
\label{staticconstraint}
\end{eqnarray}
There are two differences between this system and the class of solutions we have just found on the canonical side. First, in the Lagrangian system (\ref{extra} - \ref{staticconstraint}) there is the additional equation (\ref{extra}) and, second, the Eq.~(\ref{staticconstraint}) is different to the constraint $\mathcal{H} = 0$ (\ref{hamiltonianconstraint}) since the term $\frac{1}{2} \pi^2$ is missing. On the other hand, as a consequence of the general analysis we did previously, the system of equations (\ref{extra} - \ref{staticconstraint}) is reproduced on the Hamiltonian side (explicitly, substitute $\dot{\phi} = 0$ in Eqs.~(\ref{fromhtol1} - \ref{fromhtol2}) and redefine $n \rightarrow ( 1 + k_2 ) n$ with $k_2 \neq -1$). Therefore, all the solutions of (\ref{extra} - \ref{staticconstraint}) can be found in the Hamiltonian formalism, but there are static solutions in the Hamiltonian formalism that are not solutions of the Lagrangian formalism if the Lagrangian is the one given in (\ref{actionlagrangian}).


\section{Hamiltonian of the kinetic-conformal Ho\v{r}ava theory}
\label{sec:generalhamiltonian}
The Ho\v{r}ava theory is written in terms of the ADM variables $N$, $N_i$ and $g_{ij}$. We consider the nonprojectable version where the lapse function $N$ is a function of the time and the space. We denote by $\pi^{ij}$ the canonically conjugate momentum of $g_{ij}$ and by $P_N$ the one of $N$, which is zero. Although this is an already-solved constraint, we keep it explicit in the Hamiltonian analysis for the sake of rigorousity. The shift vector $N_i$ can be regarded as a Lagrange multiplier. An important \cite{Blas:2009qj} composed object is the FDiff-covariant vector
\begin{equation}
 a_i = \frac{ \partial_i N }{ N } \,.
\end{equation}
Curvature tensors of the spatial metric as the spatial Ricci tensor $R_{ij}$ and the spatial Ricci scalar $R$ are used in this theory. We consider only the terms that are of second order in spatial derivatives (the theory is of second order in time derivatives). The second-order terms are the dominant modes for the large-scale physics.

The complete Hamiltonian of the kinetic-conformal theory, with all the constraints added, is
\begin{equation}
H =
\int d^3x \left[
\sqrt{g} N \left(
\frac{1}{g} \pi^{ij} \pi_{ij} 
- \beta R
- \alpha  a_i a^i \right)
+ N_i \mathcal{H}^i
+ \sigma P_N
+ \mu \pi
+ A_1 \mathcal{C}_1
+ A_2 \mathcal{C}_2
\right]
\label{fullhamiltonian}
\end{equation}
$\beta$ and $\alpha$ are coupling constants. $\sigma$, $\mu$, $A_1$ and $A_2$ are the Lagrange multipliers of the constraints
\begin{eqnarray}
&& P_N = 0\,,
\label{pn}
\\
&& \pi = 0 \,,
\label{picero}
\\ &&
\frac{1}{\sqrt{g}} \mathcal{C}_1 \equiv 
\frac{2}{g} \pi^{ij} \pi_{ij}
- \gamma_1 \frac{ \nabla^2 N }{N}  = 0 \,,
\label{constraintc1}
\\ &&
\frac{1}{\sqrt{g}} \mathcal{C}_2 \equiv
\gamma_2 \frac{\nabla^2 N}{N}
- \alpha a_i a^i
- \beta R = 0 \,,
\label{constraintc2}
\end{eqnarray}
where
\begin{equation}
 \gamma_1 \equiv 2\beta - \alpha \,, \quad
 \gamma_2 \equiv \beta + \frac{3\alpha}{2} \,. 
\end{equation}
$\mathcal{H}^i$ is the momentum constraint,
\begin{equation}
\mathcal{H}^j \equiv 
- 2 \nabla_i \pi^{ij} + P_N \partial^j N = 0 \,.
\label{momentumconstraint}
\end{equation}
It is the generator of the symmetry of transformations on the spatial coordinates. This is a first-class constraint whereas the four constraints $P_N$, $\pi$, $\mathcal{C}_1$ and $\mathcal{C}_2$ are of second class.

The Hamiltonian formulation was derived in Ref.~\cite{Bellorin:2013zbp} following the standard procedure of obtainning the Hamiltonian from a Legendre trasnformation of the Lagrangian, and then demanding the preservation of the constraints. In this way, the secondary constraints that directly arise are actually
\begin{eqnarray}
 \frac{1}{ \sqrt{g} } \mathcal{H} &\equiv&
 \frac{1}{g} \pi^{ij} \pi_{ij} - \beta R + 2 \alpha \frac{\nabla^2 N}{N}
 - \alpha a_i a^i = 0 \,,
 \label{hamiltonianconstraintoriginal}
 \\
 \frac{1}{ \sqrt{g} } \mathcal{C} &\equiv&
 \frac{3}{ 2g } \pi^{ij} \pi_{ij} + \frac{\beta}{2} R
 - 2 \frac{ \nabla^2 N }{ N } + \frac{\alpha}{2} a_i a^i = 0 \,. 
 \label{cconstraintoriginal}
\end{eqnarray}
Here we have recombined the pair $\mathcal{H} = \mathcal{C} = 0$ into the pair $\mathcal{C}_1 = \mathcal{C}_2 = 0$, given in (\ref{constraintc1} - \ref{constraintc2}), since the latter have a simpler structure. We also comment that the constraints $\pi = 0$ and $\mathcal{C} = 0$ are the additional constraints that characterize the kinetic-conformal theory; they are absent out of the kinetic-conformal point, i.~e., when $\lambda \neq 1/3$ (and in that case the structure of $\mathcal{H}$ is different). 

By applying Dirac's procedure for the time preservation of the constraints in time, we obtain that the preservation of the four second-class constraints (\ref{pn} - \ref{constraintc2}) leads to the following four equations for the Lagrange multipliers,
\begin{eqnarray}
 0 &=& 
 \nabla^2 B 
 + 2 \alpha a^k \nabla_k \left( \frac{ A_2 }{ N } \right)
 + \frac{ \gamma_1 }{ 2 } \frac{ \nabla^2 N }{ N^2 } ( 2 A_1 - A_2 )
 \,,
 \label{a2equation}
 \\
 0 &=& 
   4 \beta \frac{ \nabla^2 A_2 }{ N } 
 - a_k \nabla^k B
 - \frac{ \nabla^2 N }{ N^2 } \left( 
     3 \gamma_1 A_1 + \gamma_2 A_2 \right) \,,
 \label{preservationpi}
 \\
 0 &=& 
 \gamma_1 \frac{\sqrt{g}}{N} 
   \left( \nabla^2 \sigma - \frac{ \nabla^2 N }{ N } \sigma \right)
 + \left( \frac{3}{\sqrt{g}} \pi^{ij} \pi_{ij}  
        + \frac{ \gamma_1 }{ 2 } \sqrt{g} a_k a^k \right) \mu
 \nonumber \\ &&
 - \left\{ \mathcal{C}_1 , 
     \left< ( N + 2A_1 ) \frac{\pi^{ij} \pi_{ij}}{\sqrt{g}} 
       - \sqrt{g} ( N + A_2 ) \left( \beta R + \alpha a_k a^k \right)
            + \sqrt{g} B \nabla^2 N \right> \right\} \,,
  \nonumber \\ &&
  \label{preservationc1}
  \\
  0 &=& 
  \gamma_2 \frac{\sqrt{g}}{N} 
  \left( \nabla^2 \sigma - \frac{ \nabla^2 N }{ N } \sigma \right)
  + 2 \alpha \frac{\sqrt{g}}{N} 
    a^k \left( \partial_k \sigma + a_k \sigma \right)
   + 2 \beta \sqrt{g} \nabla^2 \mu
  \nonumber \\ &&
  - \frac{ \sqrt{g} }{2} \left( \beta R 
         + ( \alpha - \gamma_2 ) a_k a^k \right) \mu
  + \left\{ \mathcal{C}_2 , 
    \left< ( N + 2 A_1 ) \frac{\pi^{ij} \pi_{ij}}{\sqrt{g}} \right> \right\}
  \,,
  \label{preservationc2}
\end{eqnarray}
where
\begin{equation} 
 B \equiv
 - \frac{1}{N} \left( \gamma_1 A_1 - \gamma_2 A_2 \right) \,.
\end{equation}
The preservation of the momentum constraint $\mathcal{H}^i = 0$ is ensured on the constrained phase space since it is a first-class constraint

To obtain the Hamiltonian evolution equations we take variations of the Hamiltonian (\ref{fullhamiltonian}) with respect to all the canonical variables. Taking variations with respect to $N$ reproduces exactly the Eq.~(\ref{a2equation}), since this equation was obtained by imposing $\dot{P}_N = 0$. Variations with respect to $\pi^{ij}$, $g_{ij}$ and $P_N$ yield, respectively, the equations
\begin{eqnarray}
 \dot{g}_{ij} &=&
 \frac{2}{\sqrt{g}} ( N + 2 A_1 ) \pi_{ij} + \mu g_{ij} \,,
 \label{gpunto}
 \\
 \frac{\dot{\pi}^{ij}}{\sqrt{g}} &=&
 - \frac{1}{g}( N + 2 A_1 ) 
     ( 2 \pi^{ik} \pi_k{}^j - \frac{1}{2} g^{ij} \pi^{kl} \pi_{kl} )
 - \mu \frac{\pi^{ij}}{\sqrt{g}}
 - \nabla^{(i} N \nabla^{j)} B + \frac{1}{2} g^{ij} \nabla_k N \nabla^k B
 \nonumber
 \\ &&
 + \left[ \beta ( \nabla^{ij} - g^{ij} \nabla^2 )
        - \beta ( R^{ij} - \frac{1}{2} g^{ij} R )
        - \alpha ( a^i a^j - g^{ij} a_k a^k ) \right] ( N + A_2 )
 \nonumber
 \\ &&
 \label{pipunto}
 \\
  \dot{N} &=& \sigma \,.
  \label{dotn}
\end{eqnarray}
In these equations we have fixed the gauge condition $N_i = 0$. 

To obtain a solution of the theory in the Hamiltonian formalism we may adopt two different, but equivalent, schemes. The first scheme is to pose an initial-data problem. The initial data on the canonical variables is subject to the constraints (\ref{pn} - \ref{constraintc2}) and (\ref{momentumconstraint}) at the initial time. The equations (\ref{a2equation} - \ref{preservationc2}) must be solved for the Lagrange multipliers for all time, since they ensure that the constraints are satisfied at all times. In this scheme the evolution equations (\ref{gpunto} - \ref{dotn}) are interpretated as the formulas giving the evolution of the canonical variables in a consistent way, starting from the initial data.

The second scheme is an ``all-time solution" problem, on which all the field equations that determine the stationary points of the action are solved simultaneosly for all time. The field equations are obtained by taking variations of the canonical action with respect to all the variables on which it depends: the canonical variables and the Lagrange multipliers. Then, a solution is given when all these equations are solved simultaneosly for all time. In this scheme the field equations that need to be simultaneously solved for all times are the constraints (\ref{pn} - \ref{constraintc2}) and (\ref{momentumconstraint}), the evolution equations (\ref{gpunto} - \ref{dotn}), and the Eq.~(\ref{a2equation}) since this equation has a dual role: it is both a condition for the preservation of constraints and a field equation for a stationary point. The two schemes are equivalent.

\section{Special Hamiltonian solutions}
\label{sec:vacuumstatic}
\subsection{Existence of special Hamiltonian solutions}
Here and in the next section we analyze configurations for which the Legendre transformation cannot be inverted. In this section we restrict the analysis to the purely gravitational theory, without coupling to matter sources. The corresponding Hamiltonian formalism was given in the previous section.

As in the toy model, the key for the special solutions is the vanishing of the coefficient of $\pi^{ij}$ in the evolution Eq.~(\ref{gpunto}). Let us analyze first, using the scheme of a problem of inital data, the existence of these special solutions, which are characterized by
\begin{equation}
 A_1 = - \frac{1}{2} N \,.
\end{equation}
We start by analyzing the equations for the Lagrange multipliers. We require that also $A_2$ is proportional to $N$,
\begin{equation}
 A_2 = k_2 N \,.
\end{equation}
With these settings for $A_{1,2}$ the variable $B$ becomes constant, and we obtain that the two Eqs.~(\ref{a2equation}) and (\ref{preservationpi}) are completely solved if $k_2 = -1$, i.~e., $A_{1,2}$ are given in terms of $N$ by
\begin{equation}
 N + 2 A_1 = N + A_2 = 0 \,.
 \label{multipliers}
\end{equation}
Next, it is easy to evaluate the Eqs.~(\ref{preservationc1} - \ref{preservationc2}) under (\ref{multipliers}) since in these equations the factors $N + 2 A_1$ and $N + A_2$ inside the brackets remain unaltered after computing the brackets ($\mathcal{C}_1$ and $\mathcal{C}_2$ do not depend on $P_N$). In addition, since $B$ becomes constant, the term $\sqrt{g} B \nabla^2 N$ in (\ref{preservationc1}) becomes a total derivative that does not contribute to the bracket. Therefore, the whole terms vanish after (\ref{multipliers}) is imposed. The Eqs.~(\ref{preservationc1} - \ref{preservationc2}) result
\begin{eqnarray}
 0 &=& 
 \frac{ \gamma_1 }{N} 
 \left( \nabla^2 - \frac{ \nabla^2 N }{ N } \right) \sigma
 + \left( \frac{ 3 }{ g } \pi^{ij} \pi_{ij}
      + \frac{ \gamma_1 }{ 2 } a_k a^k \right) \mu  
 \,,
 \label{preservationc1special}
 \\
 0 &=&
 \frac{ \gamma_2 }{N} 
 \left( \nabla^2 - \frac{ \nabla^2 N }{ N } \right) \sigma
 + \frac{ 2 \alpha }{N} 
 a^k \left( \partial_k + a_k \right) \sigma
 + 2 \beta \nabla^2 \mu
 - \frac{ 1 }{2} \left( \beta R + ( \alpha - \gamma_2 ) a_k a^k \right) \mu
 \,.
 \nonumber \\ &&
 \label{preservationc2special}
\end{eqnarray}
These two equations form a system of homogeneous partial differential equations for $\sigma$ and $\mu$. A particular solution, without requiring any further condition, is $\sigma = \mu = 0$. We shall use this solution on the next subsection. On more general grounds, the system (\ref{preservationc1special} - \ref{preservationc2special}) becomes elliptic if the matrix of coefficients of the terms of second order in derivatives is positive definite. This sets two conditions on the space of coupling constants, namely
\begin{equation}
 \beta > 0 \,, \quad
 \gamma_1 = 2\beta - \alpha > 0 \,.
\end{equation}
Therefore, if we assume these condition holds, there exist solutions of the Eqs.~(\ref{preservationc1special} - \ref{preservationc2special}), each one corresponding to a given boundary condition.

The initial data is subject to the constraints (\ref{pn} - \ref{constraintc2}) and (\ref{momentumconstraint}) at the initial time. In the set $\{ (g_{ij},\pi^{ij}),(N,P_N)\}$ of canonical variables there are 14 functional degrees of freedom over the $t=0$ slice. The set of constraints (\ref{pn} - \ref{constraintc2}) and (\ref{momentumconstraint}) plus the gauge fixing condition for the symmetry of spatial diffeomorphisms fix 10 of these degrees of freedom. Thus, there remain 4 free degrees of freedom on the initial data. This corresponds to two propagating physical modes.

The equations that govern the evolution of the initial data, Eqs.~(\ref{gpunto} - \ref{dotn}), get greatly simplified over these special configurations. Indeed, in these equations there are several ocurrences of the factors $N + 2A_1$, $N + A_2$, and derivatives of $B$. All of them become zero on these special solutions. Therefore, the equations for the evolution of the canonical variables take the simple form
\begin{eqnarray}
 \dot{g}_{ij} &=&
 \mu g_{ij} \,,
 \label{gpuntospecial}
\\
 \dot{\pi}^{ij} &=&
 - \mu \pi^{ij} \,,
\label{pipuntospecial}
\\
\dot{N} &=& \sigma \,.
\label{dotnspecial}
\end{eqnarray}

Therefore, we have that a class of interesting special solutions exists. They represent a flow of the three-dimensional metric on a given topology. These solutions are characterized by $2A_1 = A_2 = -N$. The remaining Lagrange multipliers $\sigma$ and $\mu$ are determined by the equations (\ref{preservationc1special} - \ref{preservationc2special}), and the initial data on the canonical variables must solve the constraints (\ref{pn} - \ref{constraintc2}) and (\ref{momentumconstraint}). The evolution equations are (\ref{gpuntospecial} - \ref{dotnspecial}). Notice that the evolution flows of $g_{ij}$ and $\pi^{ij}$ are coupled each other due to the factor $\mu$. This class of solutions cannot be obtained from the original Lagrangian field equations since there is no direct relation between $\dot{g}_{ij}$ and $\pi^{ij}$.


\subsection{Example: static solutions}
Here we pursue a concrete example of solutions of the equations that we have previously found. We start with the equations for the Lagrange multipliers $\sigma$ and $\mu$, Eqs.~(\ref{preservationc1special} - \ref{preservationc2special}). Since these equations are homogeneous, an obvious solution is 
\begin{equation}
 \sigma = \mu = 0 \,.
\end{equation}
With this the right-hand sides of Eqs.~(\ref{gpuntospecial} - \ref{dotnspecial}) vanish, which implies that the configurations have static canonical fields, $\dot{g}_{ij} = \dot{\pi}^{ij} = \dot{N} = 0$. Therefore, the initial data is the whole solution since it is static.

After imposing $P_N = 0$ explicitly, we have that the static solutions must satisfy the constraints (\ref{picero} - \ref{constraintc2}) and (\ref{momentumconstraint}). It is convenient to use the constraint $\mathcal{C}_2$ to solve $N$. To this end the following change of variables is useful,
\begin{equation}
 N = W^\gamma \,, \quad
 \gamma \equiv \frac{ 2 \gamma_2 }{ 2 \beta + \alpha } \,.
 \label{nw}
\end{equation}
With this change the constraint $\mathcal{C}_2$ given in (\ref{constraintc2}) becomes
\begin{equation}
   \nabla^2 W = \chi_1 R W  \,,
   \quad
   \chi_1 \equiv \frac{  \beta ( 2 \beta + \alpha ) }{ 2 \gamma_2^2 } \,,
 \label{eqw}
\end{equation}
which we regard as an equation for $W$. We now use (\ref{eqw}) in the constraint $\mathcal{C}_1$ given in (\ref{constraintc1}), obtaining
\begin{equation}
 \frac{1}{g} \pi^{ij} \pi_{ij} =
   2 \chi_2 \frac{ \partial_i W \partial^i W }{ W^2 }
 + 2 \chi_3 R \,,
 \label{eqpisquare}
\end{equation}
where
\begin{equation}
  \chi_2 \equiv 
  \frac{ \alpha \gamma_1 \gamma_2 }
  { ( 2\beta + \alpha )^2 }  \,,
  \quad
  \chi_3 \equiv 
  \frac{ \beta ( 2 \beta + \alpha ) }{ 4 \gamma_2 } \,.
\end{equation}
The constraints (\ref{picero}), (\ref{momentumconstraint}) and (\ref{eqpisquare}) can be regarded as a set of conditions on $\pi^{ij}$.

To further advance, we assume that we are in a region on which the static spatial metric $g_{ij}$ is a flat metric. Then the Eq.~(\ref{eqw}) takes the form
\begin{equation}
  \nabla^2 W = 0 \,,
  \label{harmonicw}
\end{equation}
where now $\nabla^2$ is the flat Laplacian. Therefore, any harmonic function $W$ on the given region satisfying some boundary condition solves the constraint (\ref{harmonicw}). There remain the constraints on $\pi^{ij}$, which we sumarize here,
\begin{eqnarray}
 \pi &=& 0 \,,
 \label{picero2}
 \\
  \nabla_i \pi^{ij} &=& 0 \,,
  \label{momentumconstraint2}
 \\
   \frac{1}{g} \pi^{ij} \pi_{ij} &=&
   2 \chi_2 \frac{ \partial_i W \partial^i W }{ W^2 } \,.
   \label{pisquare}
\end{eqnarray}
The consistency of Eq.~(\ref{pisquare}), for any configuration with  $\pi^{ij} , \partial_i W \neq 0$, requires $\chi_2 \geq 0$. This is a condition on the coupling constants $\alpha$ and $\beta$.

We may give a class of static solutions of the system (\ref{harmonicw} - \ref{pisquare}) with the flat metric in the following way. We implement Cartesian coordinates on the spatial slices. The solutions start with the assumption that one is given with a harmonic function $W$ that can be expressed in separate variables,
\begin{equation}
 W = X^1(x^1) X^2(x^2) X^3(x^3) \,.
 \label{separated}
\end{equation}
For the momentum field we assume that all its diagonal components vanish, $\pi^{11} = \pi^{22} = \pi^{33} = 0$, such that the constraint (\ref{picero2}) is automatically solved. The momentum constraint (\ref{momentumconstraint2}) is completely solved if the off-diagonal components have the following dependence on the coordinates:
\begin{equation}
 \pi^{12}(x^3) \,, \quad \pi^{13}(x^2) \,, \quad \pi^{23}(x^1) \,.
\end{equation}
Finally, the constraint (\ref{pisquare}) is solved if these components of the momentum are given in terms of the components of the harmonic function in the form
\begin{equation}
 \pi^{12} =  \frac{\sqrt{\chi_2}}{X^3} \frac{dX^3}{dx^3} \,,
 \quad
 \pi^{13} =  \frac{\sqrt{\chi_2}}{X^2} \frac{dX^2}{dx^2} \,,
 \quad
 \pi^{23} =  \frac{\sqrt{\chi_2}}{X^1} \frac{dX^1}{dx^1} \,.
 \label{solutionmomentum}
\end{equation}

Summarizing, the new vacuum solutions we have presented are composed of the following ingredients: all canonical fields and Lagrange multipliers are static. The spatial metric is the flat one, in Cartesian coordinates $g_{ij} = \delta_{ij}$. With this flat metric in Cartesian coordinates pick up a harmonic function $W$ that can be expressed in separate variables as (\ref{separated}). With the components of the harmonic function build up the off-diagonal components of the conjugated momentum according to (\ref{solutionmomentum}). Put the diagonal components equal to zero.  The lapse function $N$ is given in terms of the harmonic function $W$ by Eq.~(\ref{nw}). Finally, the Lagrange multipliers are $\sigma = \mu = 0$ and $A_2 = 2 A_1 = -N$.

\section{Coupling to sources: homogeneity and isotropy}
\label{sec:cosmosolution}
The ways on which matter sources can be coupled to the Ho\v{r}ava gravity are in principle different to GR, since the Ho\v{r}ava gravity is a nonrelativistic theory. Our proposal here is not to undertake this discussion, but to find a suitable scenario under which concrete special solutions can be found on the Hamiltonian formalism. In particular we ask ourselves whether solutions coupled to homogeneous and isotropic sources can be found. With this aim, and paralellizing the standard approach of GR, we consider a perfect fluid that has a relativistic energy-momentum tensor,
\begin{equation}
T_{\mu\nu} =
\rho u_\mu u_\nu
+ P (g_{\mu\nu} + u_\mu u_\nu) \,.
\label{perfectfluid}
\end{equation}
We also consider that the perfect fluid is at rest in the chosen reference frame, which implies that in this frame the four-velocity takes the form
\begin{equation}
 u^\mu = ( N^{-1} , 0 , 0 , 0 ) \,,
\end{equation}
such that the constraint $u_\mu u^\mu = -1$ is satisfied. 

A words of caution about this coupling are necessary. Since this energy-momentum tensor is going to be coupled to the nonrelativistic gravity, it cannot be taken for granted that the gravitational field equations do not induce nonrelativistic effects on the behavior of the source. One of such nonrelativistic effects is that the field equations in the Hamiltonian formalism do not imply the relativistic energy-momentum conservation condition $\nabla_\mu T^{\mu\nu} = 0$. The corresponding conservation law that adopts the source in this theory can, in principle, be deduced from the Hamiltonian field equations (or directly on a given solution).

Since in the case of the coupling to a source we want to find concrete solutions explicitly, we work on the approach of the all-time solution, where the field equations for the stationary point need to be solved. We substitute the constraint $P_N = 0$ explicitly throughout the analysis since it is unnecessary for our task here. In this case the Eq.~(\ref{dotn}), which gets no modification by the coupling to the source, can be simply interpreted as the equation determining $\sigma$, which becomes an irrelevant variable. The Eqs.~(\ref{picero}), (\ref{momentumconstraint}) and (\ref{a2equation}) receive no contribution from the source, whereas the constraints $\mathcal{C}_1$ and $\mathcal{C}_2$ and the evolution equations (\ref{gpunto}) and (\ref{pipunto}) get modified by it. We summarize here the resulting system of field equations,
\begin{eqnarray}
\pi &=& 0 \,,
\label{picerosourced}
\\ 
\nabla_i \pi^{ij} &=& 0 \,,
\label{momentumconstraintsourced}
\\
\frac{1}{\sqrt{g}} \mathcal{C}_1 &\equiv& 
\frac{2}{g} \pi^{ij} \pi_{ij}
- \gamma_1 \frac{ \nabla^2 N }{N}
+ \rho + 3 P  = 0 \,,
\label{constraintc1sourced}
\\ 
\frac{1}{\sqrt{g}} \mathcal{C}_2 &\equiv&
\gamma_2 \frac{\nabla^2 N}{N}
- \alpha a_i a^i
- \beta R 
+ \frac{3}{2} ( \rho - P ) = 0 \,,
\label{constraintc2sourced}
\\
\dot{g}_{ij} &=&
\frac{2}{\sqrt{g}} ( N + 2 A_1 ) \pi_{ij} + \mu g_{ij} \,,
\label{gpuntosourced}
\\
\frac{\dot{\pi}^{ij}}{\sqrt{g}} &=&
- \frac{1}{g}( N + 2 A_1 ) 
( 2 \pi^{ik} \pi_k{}^j - \frac{1}{2} g^{ij} \pi^{kl} \pi_{kl} )
- \mu \frac{\pi^{ij}}{\sqrt{g}}
- \nabla^{(i} N \nabla^{j)} B + \frac{1}{2} g^{ij} \nabla_k N \nabla^k B
\nonumber
\\ &&
+ \left[ \beta ( \nabla^{ij} - g^{ij} \nabla^2 )
- \beta ( R^{ij} - \frac{1}{2} g^{ij} R )
- \alpha ( a^i a^j - g^{ij} a_k a^k ) \right] ( N + A_2 )
\nonumber
\\ &&
\nonumber
\\ &&
- \frac{1}{2} g^{ij} \left[
( A_1 + \frac{3}{2} A_2 ) \rho
- ( 2 N - 3 A_1 + \frac{3}{2} A_2 ) P \right] \,,
\label{pipuntosourced}
\\
 0 &=& 
 \nabla^2 B 
 + 2 \alpha a^k \nabla_k \left( \frac{ A_2 }{ N } \right)
 + \frac{ \gamma_1 }{ 2 } \frac{ \nabla^2 N }{ N^2 } ( 2 A_1 - A_2 )
 \,.
 \label{a2equationsourced}
\end{eqnarray}

As in the previous vacuum case, we pursue special solutions for which the Hamiltonian formulation looses the relationship with the original Lagrangian. These solutions are characterized by $ 2A_1 = -N$. Since the Eq.~(\ref{a2equationsourced}) remains unaltered after the coupling to the source, adopting the same previous ansatz for the Lagrange multiplier $A_2$ seems promisory. Thus, we fix again the Lagrange multipliers $A_1$ and $A_2$ according to $2 A_1 = A_2 = -N$, such that Eq.~(\ref{a2equationsourced}) is completely solved. We consider that the source variables $\rho$ and $P$ depend only on the time coordinate. We introduce the ansatz on which the spatial metric is homogeneous and isotropic. For the sake of simplicity we only consider the flat case. In Cartesian coordinates, this ansatz is impleted by the spatial metric
\begin{equation}
ds^2_{(3)} = a^2(t) \left( (dx^1)^2 + (dx^2)^2 + (dx^3)^2 \right) \,.
\label{homogeneousisotro}
\end{equation} 
We also consider that the momentum $\pi^{ij}$ depends only on the time coordinate. We do not consider that the lapse function $N$ is a function only of time since this leads to an uninteresting restriction, as we shall see. With these settings the momentum constraint (\ref{momentumconstraintsourced}) is automatically solved.

Equation (\ref{gpuntosourced}) reduces to
\begin{equation}
 2 a \dot{a} \delta_{ij} =
 \mu a^2 \delta_{ij} \,,
\label{gpuntosourcedred}
\end{equation}
and its solution is
\begin{equation}
\mu = \frac{2 \dot{a}}{a} \,.
\label{solutionmu}
\end{equation}
Equation (\ref{pipuntosourced}) becomes
\begin{equation}
\dot{\pi}^{ij} + \frac{ 2 \dot{a} }{ a } \pi^{ij} =
\delta^{ij} a N ( \rho + P ) \,.
\label{presolutionpi}
\end{equation}
Since constraint (\ref{picerosourced}) demands that $\pi^{ij}$ is traceless, from (\ref{presolutionpi}) we extract the equation of state
\begin{equation}
P = - \rho \,.
\label{equationofstate}
\end{equation}
In turn, this implies that the right hand side of Eq.~(\ref{presolutionpi}) is equal to zero. This equation can be directly integrated, its solution is
\begin{equation}
\pi^{ij} = \frac{ m^{ij} }{ a^2 } \,,
\label{solutionpi}
\end{equation}
where $m^{ij}$ is a symmetric traceless constant matrix, such that the constraint (\ref{picerosourced}) is automatically solved.

There remain the constraints $\mathcal{C}_1$ and $\mathcal{C}_2$, given in (\ref{constraintc1sourced}) and (\ref{constraintc2sourced}), to be solved. Under the conditions we are considering, these constraints take, respectively, the form
\begin{eqnarray}
\frac{ 2 m^2 }{ a^4 } 
- ( 2\beta - \alpha ) \frac{ \partial^2 N }{N} &=&
2 a^2 \rho \,,
\label{constraintc1ansatz}
\\
( \beta + \frac{3\alpha}{2} ) \frac{\partial^2 N}{N}
- \alpha \frac{ \partial_i N \partial_i N }{ N^2 } &=& 
- 3 a^2 \rho  \,,
\label{constraintc2ansatz}
\end{eqnarray}
where $m^2 \equiv m^{ij} m^{ij}$ and $\partial^2$ is the flat Laplacian in Cartesian coordinates, $\partial^2 = \partial_i \partial_i$. Notice that if the lapse function $N$ was a function only of time, then the constraint (\ref{constraintc2ansatz}) would imply $a^2 \rho = 0$. Because of this we relax the functional form $N$, allowing a spatial dependence for it in such a way that these two constraints can be solved consistently. We introduce the following anstaz for $N$,
\begin{equation}
N = \exp\left( f_0(t) + f_1(t) x^1 + f_2(t) x^2 + f_3(t) x^3 \right) \,,
\label{solutionn}
\end{equation}
where $f_{0,1,2,3}(t)$ are arbitrary functions of time. Then the Eq.~(\ref{constraintc2ansatz}) reduces to
\begin{equation}
f_i f_i = - \frac{ 6 a^2 \rho }{ 2 \beta + \alpha } \,,
\label{conditionfsquare}
\end{equation}
where $f_i f_i = f_1^2 + f_2^2 + f_3^2$. Consistency of this equation requires that the coupling constants satisfy the bound $2\beta + \alpha < 0$ (assuming $\rho > 0$). By using Eq.~(\ref{conditionfsquare}) in Eq.~(\ref{constraintc1ansatz}), we obtain that (\ref{constraintc1ansatz}) becomes an algebraic equation for $a$ and $\rho$, whose solution is
\begin{equation}
\rho = k a^{-6} \,, \quad
k \equiv \frac{ | 2\beta + \alpha | m^2 }{ 4 (\beta - \alpha) } \,.
\label{rho}
\end{equation}
(If we assume $\beta > 0$ and $2\beta + \alpha < 0$, then $\beta - \alpha > 0$). 

In summary, the solution has the homogeneous and isotropic spatial slices whose metric is given in (\ref{homogeneousisotro}). $a$ and $\rho$ are related by Eq.~(\ref{rho}) and the solution requires the equation of state (\ref{equationofstate}). Note that, although $P = - \rho$, the density and the pressure are not constant in this solution. This is admissible since, as we pointed out above, the fluid does not satisfy the relativistic energy-momentum conservation condition. The conjugated momentum $\pi^{ij}$ is given in terms of $a$ by (\ref{solutionpi}), where the constant matrix $m^{ij}$ is symmetric and traceless, but otherwise arbitrary. The lapse funtion $N$ is given by Eq.~(\ref{solutionn}), where the functions $f_{1,2,3}$ are subject to the condition (\ref{conditionfsquare}). Upon the relation (\ref{rho}), a solution of Eq.~(\ref{conditionfsquare}) is
\begin{equation}
f_1 = f_2 = f_3 = \sqrt{\frac{ 2 k }{ | 2\beta + \alpha | }} \, a^{-2} \,.
\end{equation}
$f_0$ is left undetermined. Finally, the Lagrange multipliers are fixed by 
(\ref{solutionmu}), $\sigma = \dot{N}$, and $2 A_1 = A_2 = -N$. The solution is valid in the region $2\beta + \alpha < 0$ of the space of coupling constants.

\section{Comparison with the Lagrangian}
\label{sec:lagrangian}
The action expressed in terms of the original Lagrangian of the nonprojectable Ho\v{r}ava theory, up to second order in derivatives, is \cite{Horava:2009uw,Blas:2009qj}
\begin{equation}
S =  \int dt d^3x \sqrt{g} N \left (
G^{ijkl} K_{ij} K_{kl} + \beta R + \alpha a_i a^i \right) \,,
\end{equation}
where
\begin{eqnarray}
&& G^{ijkl} = 
\frac{1}{2} ( g^{ik} g^{jl} + g^{il} g^{jk} ) - \lambda g^{ij} g^{kl} \,,
\\
&& K_{ij} = \frac{1}{2N} ( \dot{g}_{ij} - 2 \nabla_{(i} N_{j)} ) \,.
\end{eqnarray}
The kinetic-conformal formulation we study here is based on setting $\lambda = 1/3$. We couple again the gravitational theory to the perfect fluid with the energy-momentum tensor (\ref{perfectfluid}).
The equations of motion are obtained by taking variations with respect to $g_{ij}$, $N$ and $N_i$. This yields the equations
\begin{eqnarray}
\frac{1}{\sqrt{g}} \frac{\partial}{\partial t} 
( \sqrt{g} G^{ijkl} K_{kl} )
+ 2 G^{klm(i|} \nabla_k ( K_{lm} N^{|j)} )
- G^{ijkl} \nabla_m (  K_{kl} N^m )
& &\nonumber \\
+ 2 N ( K^{ik} K_k{}^j - \lambda K K^{ij} )
- \frac{1}{2} N g^{ij} G^{klmn} K_{kl} K_{mn}  
& & \nonumber \\ 
+ \beta N (R^{ij} - \frac{1}{2} g^{ij} R )
- \beta ( \nabla^i \nabla^j N - g^{ij} \nabla^2 N )
&& \nonumber \\
+ \alpha N^{-1} ( \nabla^i N \nabla^j N 
- \frac{1}{2} g^{ij} \nabla_k N \nabla^k N )
& = & g^{ij} N P \,,
\nonumber \\
\label{eomgij}
\\
G^{ijkl} K_{ij} K_{kl} - \beta R 
+ 2 \alpha N^{-2} ( N \nabla^2 N - \frac{1}{2} \nabla_i N \nabla^i N ) 
& = & 2\rho \,,
\label{eomn}
\\
G^{ijkl} \nabla_j K_{kl} &=& 0 \,.
\label{eomni}
\end{eqnarray}

To perform the comparison with the Hamiltonian formalism, it is illustrative to substitute in the field equations (\ref{eomgij} - \ref{eomni}) the $N_i= 0$ condition and the final form of the spatial metric $g_{ij}$, regarding the resulting equations as conditions for the lapse function $N$ (and the source).

\begin{enumerate}
	
	\item 
	{\bf Static vacuum solutions}\\
	The solutions we found in section \ref{sec:vacuumstatic} are vacuum solutions, $P = \rho = 0$, have static fields and a flat spatial metric. Under these conditions the Eq.~(\ref{eomni}) is completely solved whereas the Eq.~(\ref{eomn}) and the trace of Eq.~(\ref{eomgij}) become, respectively,
	\begin{eqnarray}
	N \nabla^2 N - \frac{1}{2} \nabla_i N \nabla^i N  
	&=& 0 \,,
	\\
	2 \beta N \nabla^2 N - \frac{\alpha}{2} \nabla_i N \nabla^i N &=& 0 \,.
	\end{eqnarray}
	For $2\beta - \alpha \neq 0$, which was assumed in section \ref{sec:vacuumstatic}, these equations imply $\nabla_i N \nabla^i N = 0$. The only solution for static $N$ is $N = \mbox{constant}$. This contradicts the solution for $N$ given in Eqs.~(\ref{nw}) and (\ref{harmonicw}) if we consider a nonconstant harmonic function $W$. Therefore, the static vacuum solutions of section \ref{sec:vacuumstatic} are not included in the space of solutions of the Lagrangian equations (\ref{eomgij} - \ref{eomni}).
	
	\item
	{\bf Sourced homogeneous and isotropic solution}\\
	By substituting $g_{ij} = a^2(t) \delta_{ij}$ in Eq.~(\ref{eomgij}) we get that it takes the form
	\begin{equation}
	\beta N^{-1} ( \partial_{ij} N - \delta_{ij} \partial^2 N )
	- \alpha N^{-2} ( 
	\partial_i N \partial_j N - \frac{1}{2} \delta_{ij} \partial_k N \partial_k N ) = - \delta_{ij} a^2 P
	\label{noway}
	\end{equation}
	If we now substitute the ansatz (\ref{solutionn}) in this equation and examine its off-diagonal components, we get that they become
	\begin{equation}
	( \beta - \alpha ) f_i f_j = 0 \,, \quad i \neq j \,.
	\end{equation}
	For $\beta \neq \alpha$, which is implicit in the solutions of section \ref{sec:cosmosolution}, this equation implies that at least two of the three functions $f_{1,2,3}$ are equal to zero. Suppose that $f_{\underline{i}}$ denotes the only one of them that remains nonzero. Then, it is easy to check that the diagonal components of the Eq.~(\ref{noway}) yield two equations for $f_{\underline{i}}$ and $a^2 P$, whose unique solution is $f_{\underline{i}} = a^2 P = 0$ (another condition on the coupling constants, that is implicit in the analysis of section \ref{sec:cosmosolution}, arises). Therefore all $f_i$ are zero, $N$ is at most a function of time. Equation (\ref{eomn}) then implies $\rho = 0$. These results and $P= 0$ contradicts the solution we showed in section \ref{sec:cosmosolution}. 
	
\end{enumerate}


\section*{Conclusions}
We have shown how in the kinetic-conformal formulation of the nonprojectable Ho\v{r}ava theory the space of solutions of the Hamiltonian formulation is bigger than the corresponding space in the original Lagrangian formulation. This inmediatly brings our attention to the fact that the solutions found in the Lagrangian formulation are incomplete, there are more solutions. The Hamiltonian formulation is the appropiated way to address the quantization of the theory, and the Ho\v{r}ava theory was proposed as a consistent theory of quantum gravity.

A key role is played by the Lagrange multipliers associated to the secondary constraints. These contraints must be added to the Hamiltonian that is taken as the basis for the quantization. In particular, there are configurations of these Lagrange multipliers for which the Legendre transformation cannot be inverted. By working on these special cases we have found new solutions in the Hamiltonian formalism that cannot be found in the Lagrangian formalism.

To illustrate the mechanism with specific examples we have presented a large class of evolving solutions governed by the flow of a three-dimensional metric, new static vacuum solutions and a new solution with a source. The solution with the source, which specifically is a homogeneous and isotropic perfect fluid with a relativistic energy-momentum tensor, has the interesting feature of possessing a homogeneous and isotropic spatial metric with nontrivial (time dependent) scale factor. Although the coupling to perfect fluids in Ho\v{r}ava gravity does not need to be relativistic, we have taken the relativistic energy-momentum tensor as a test example.

As we have commented, in the Lagrangian formulation of the kinetic-conformal theory the homogenous and isotropic spacetime configurations coupled to perfect fluids have the uninteresting restriction that the density and the pressure vanish. Our results indicate how more homogeneous and isotropic configurations arise when, first, the solutions are studied in the Hamiltonian formalism and, second, the condition of homogeneity and isotropy is restricted to the spatial metric, whereas the lapse function is allowed to have a more general dependence on the space. The ADM variables are the natural variables of the Ho\v{r}ava theory but the interpretation of the geometry is different to GR since the fundamental geometric structure is the foliation, not the spacetime metric. The underlying symmetry is not the general covariance, but the FDiff.

Thus, the existence of the new solutions of the Hamiltonian field equations testifies that there are more solutions than in the original Lagrangian field equations. For the particular case of the homogeneous and isotropic configurations, our conclusion is that the search does not end with the restriction arising in the original Lagrangian formalism.


\section*{Acknowledgments}
A. R. is partially supported by grant Fondecyt No.~1161192, Chile.



\end{document}